\documentclass[fleqn,twoside]{article}
\usepackage{espcrc2}
\usepackage{graphicx}

\usepackage{latexsym}

\newcommand{\tmpred}{red }
\newcommand{\tmpblue}{blue }
\newcommand{\tmpgreen}{green }

\newcommand{\tmpmagenta}{magenta }

\def\BE{\begin{displaymath}}
\def\EE{\end{displaymath}}
\def\BEA{\begin{eqnarray*}}
\def\EEA{\end{eqnarray*}}
\def\BNEA{\begin{eqnarray}}
\def\ENEA{\end{eqnarray}}

\title{
The $\Omega^-$ and the strange quark mass
}
\author{
D. Toussaint
\address{
Physics Department, University of Arizona, Tucson, AZ 85721, USA
}
and
C.T.H. Davies
\address{
Department of Physics and Astronomy, University of Glasgow, Glasgow G12 8QQ, UK
}
}

\begin{document}

\begin{abstract}
$\Omega^-$ correlators have been calculated on the MILC collaboration's archive
of three flavor improved staggered quark lattices.
The $\Omega^-$ is stable under strong interactions (140 MeV
below threshold).  It provides a valuable consistency check
on a combination of strange quark mass and lattice scale
determination from other quantities.
Alternatively, the $\Omega^-$ mass could be used to fix the strange
quark mass, which gives a check on computations
of the strange quark mass based on the kaon mass.
\end{abstract}
\maketitle

Although the most timely results of lattice QCD are otherwise
unknown hadronic matrix elements needed for determination of
CKM matrix elements, precise computations of experimentally
well known quantities are essential tests of our methods.
A small number of such quantities have been computed using
three flavors of improved staggered sea quarks, showing agreement
with experiment at the few percent level\cite{PRL}.
The mass of the $\Omega^-$ baryon is another such quantity, since
the particle is well below the threshold for strong decays and
its mass is precisely known.   From the theoretical side, the 
extrapolation to the physical light quark mass is much better controlled
for this particle than for baryons containing light valence quarks.
The one complication is that the $\Omega^-$ mass is very sensitive
to the mass of the strange quark --- indeed, we could use it as a way
of fixing the strange quark mass.
Recently the HPQCD, MILC, and UKQCD collaborations have determined
the physical strange quark mass in these lattice simulations from
a study of pseudoscalar meson properties, which, roughly speaking,
amounts to fixing the strange quark mass by tuning the kaon mass\cite{QUARKMASS04}.
With the strange quark mass fixed, the computation of the $\Omega^-$
mass becomes another test of our lattice methods.

\begin{figure}[tb]
\scalebox{0.40}{\includegraphics{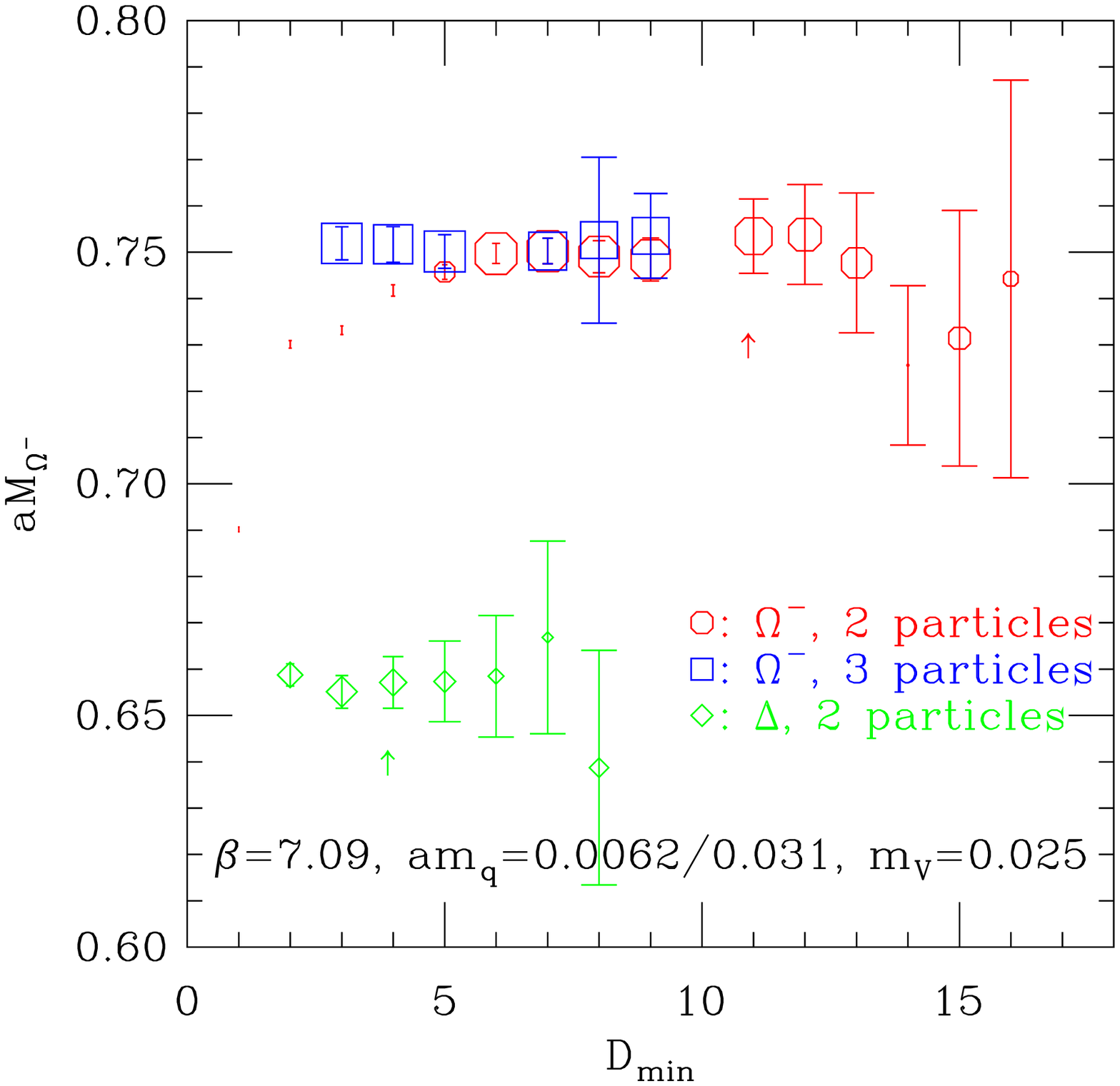}}
\vspace{-0.4in}
\caption{
Omega mass fits versus minimum distance.
This is the
partially quenched $\Omega^-$, with valence mass close to the
correct strange quark mass.  The \tmpred octagons are fits with one
state of each parity; \tmpblue squares include an (unconstrained)
excited $\Omega$ state.   The \tmpgreen diamonds are $\Delta$ fits.
Symbol sizes are proportional to the confidence level, with the
size of the legend corresponding to 50\%.
}
\label{omegaprop_fig}
\end{figure}

\begin{figure}[htb]
\scalebox{0.40}{\includegraphics{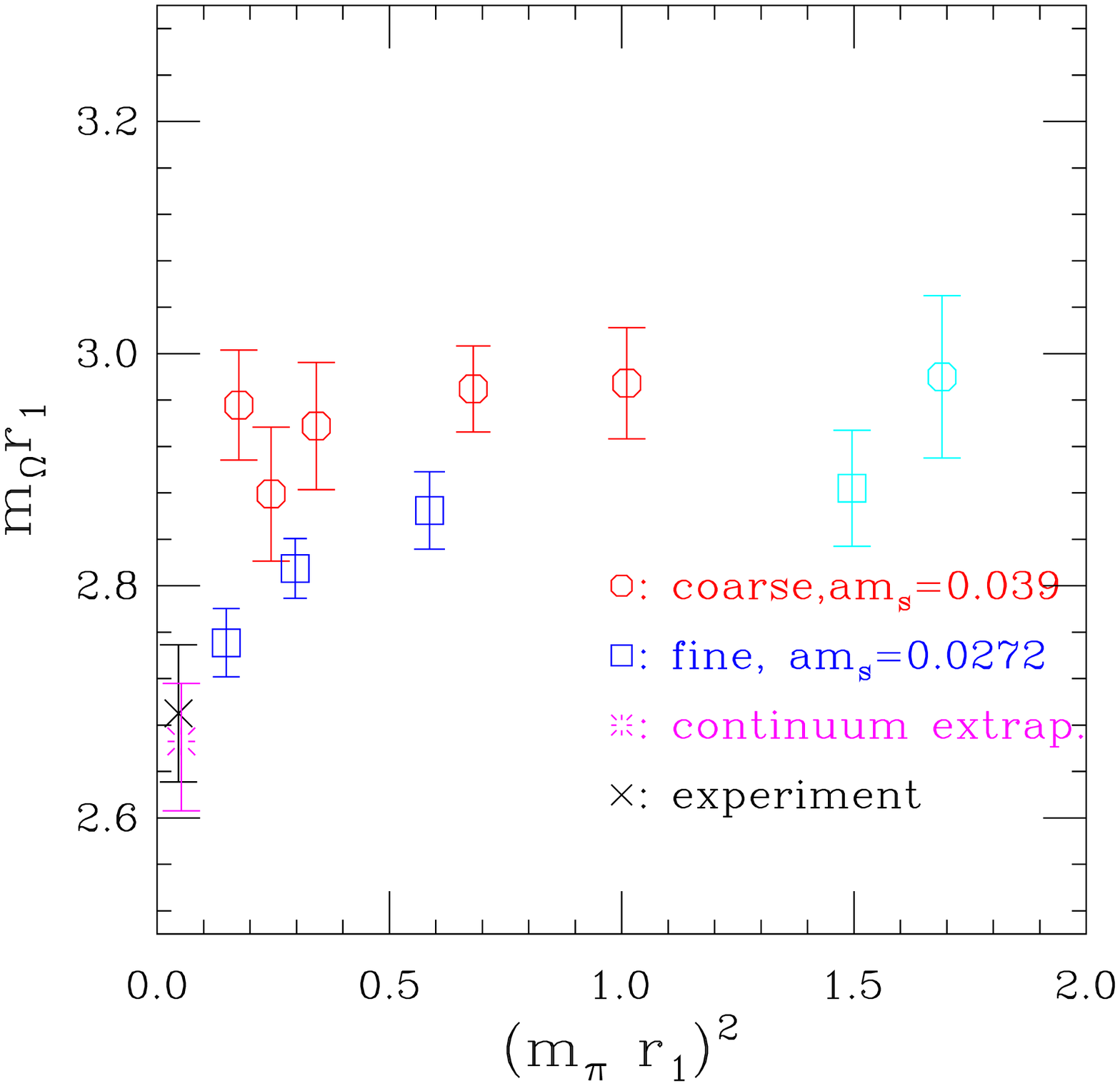}}
\vspace{-0.4in}
\caption{
$\Omega^-$ mass in units of $r_1$, with the valence
strange quark mass interpolated to
the strange quark mass determined from $M_K$,
with the continuum/chiral extrapolation and the experimental
value.
}
\label{momega_r1_fig}
\end{figure}


The NLO chiral corrections to the decuplet baryon masses
come from loop diagrams where the baryon splits into
a pseudoscalar meson and either a decuplet or octet baryon.
Isospin and strangeness conservation and the absence of an
octet baryon with $S=-3$ insure that the diagrams containing
a pion do not contribute to the $\Omega^-$ mass.
Diagrams with a kaon are present, but
the kaon mass does not vanish as $m_{u,d}\rightarrow 0$.
Higher order corrections will of course be present, but a linear
(or polynomial) extrapolation in light quark mass is much better
for the $\Omega^-$ than for other baryons.

To isolate the decuplet baryons,
a nonlocal operator is needed with Kogut-Susskind quarks.  We use
an operator from Ref.~\cite{DELTA_OPS}, following the
implementation of Ref.\cite{GGKS}, with code written by C.~DeTar\cite{DETARCODE}.
While the $\Omega^-$ correlator is statistically significant to larger
distances than the $\Delta$ correlator because of its heavier
valence quarks, the decuplet baryons' larger masses make them
relatively noisier and more difficult to fit than the octet
baryons.


\begin{table*}[htb]
\caption{
\label{fit_table}
$\Omega^-$ masses in lattice units.   Valence masses are the dynamical
strange quark mass and a partially quenched $\Omega^-$ with a valence
mass close to the strange quark mass determined from $M_K$.
The remaining columns are the sea quark masses, the number of configurations
used, the fitted mass in units of the lattice spacing, the distance range
used in the fit, the $\chi^2$ and number of degrees of freedom, and the
confidence level of the fit.
}
\begin{tabular}{|llrllll|}
\hline
$am_{valence}$ & $am_{sea}$ & $N_{conf.}$ & $aM_{\Omega^-}$ & range & $\chi^2/D$ & conf. \\
\multicolumn{7}{|l|}{ $a \approx 0.12$ fm. }\\
0.05 ($\Omega$) & 0.03/0.05     & 572	& 1.168(11)     & 7--15 & 2.8/5 & 0.74 \\
0.04 ($\Omega_{PQ}$)   & 0.03/0.05     & 572	& 1.122(18)     & 7--13 & 1.8/3 & 0.61 \\
0.05 ($\Omega$) & 0.02/0.05     & 485	& 1.169(9)      & 7--14 & 0.9/4        & 0.93 \\
0.04 ($\Omega_{PQ}$)   & 0.02/0.05     & 485	& 1.125(14)     & 7--14 & 1.7/4 & 0.80 \\
0.05 ($\Omega$) & 0.01/0.05     & 659	& 1.175(16)     & 8--15 & 2.8/4 & 0.58 \\
0.04 ($\Omega_{PQ}$)   & 0.01/0.05     & 659	& 1.130(21)     & 8--14 & 1.3/3 & 0.72 \\
0.05 ($\Omega$) & 0.007/0.05    & 498	& 1.155(14)     & 8--14 & 1.1/3 & 0.79 \\
0.04 ($\Omega_{PQ}$)  & 0.007/0.05    & 498	& 1.099(22)     & 8--14 & 2.1/3 & 0.55 \\
0.05 ($\Omega$) & 0.005/0.05    & 397	& 1.178(11)     & 9--16 & 0.8/4        & 0.93 \\
0.05 ($\Omega_{PQ}$)  & 0.005/0.05    & 397	& 1.128(18)     & 9--14 & 0.1/2 & 0.93 \\

\multicolumn{7}{|l|}{ $a \approx 0.09$ fm. }\\
0.031 ($\Omega$)        & 0.0124/0.031  & 535	& 0.791(6)      & 11--22        & 2.9/8 & 0.94 \\
0.025 ($\Omega_{PQ}$) & 0.0124/0.031  & 535	& 0.761(9)      & 11--20        & 1.0/6   & 0.98 \\
0.031 ($\Omega$)        & 0.0062/0.031  & 575	& 0.785(5)      & 11--23        & 8.8/9 & 0.46 \\
0.025 ($\Omega_{PQ}$) & 0.0062/0.031  & 575	& 0.752(7)      & 11--21        & 7.1/7 & 0.42 \\
0.031 ($\Omega$)        & 0.0031/0.031  & 106	& 0.773(5)      & 12--23        & 6.0/8        & 0.65 \\
0.025 ($\Omega_{PQ}$) & 0.0031/0.031  & 106	& 0.736(8)      & 12--22        & 5.6/7 & 0.59 \\
\hline
\end{tabular}\\
\end{table*}

Correlators  for the $\Omega^-$
were calculated on the MILC collaboration three flavor
improved staggered quark lattices, with light quark masses ranging
from $m_s$ to $m_s/9$ and lattice spacings of about 0.12 fm
(``coarse'') and 0.09 fm (``fine'')\cite{MILC_SPECTRUM2}.
Table~\ref{fit_table} contains a summary of the parameter values used
and the $\Omega^-$ mass fits, in units of the lattice spacing.
Figure~\ref{omegaprop_fig} shows the fitted masses as a function
of minimum distance included in the fit for one of the data sets.
Two valence strange quark masses were used, and the results
interpolated or extrapolated to the strange quark masses
determined from pseudoscalar meson properties, $am_s=0.039$
on the coarse lattices and $0.0272$ on the fine lattices.
These masses were fit to constant plus linear in
$\alpha a^2$ and $(m_\pi r_1)^2$, with $\chi^2=3.4$ for 4 dof., and the fit was evaluated at
$a=0$ and the physical $m_\pi$.
Figure~\ref{momega_r1_fig} shows the coarse and fine lattice
$\Omega^-$ masses as a function of light quark mass, together
with the extrapolation to the continuum and chiral limits.
In this figure the \tmpred octagons are from coarse lattices
and the \tmpblue squares from fine.
The \tmpmagenta burst at the left is the chiral
and continuum extrapolation, and the cross at the
left the experimental value with an error from the
uncertainty in $r_1$.  The two right hand points 
are from runs with $m_{u,d}=m_s$ adjusted to the
desired $m_s$.  These two points were not used in the
fit for the central value, but were used in fits
estimating the effects of higher order terms in $a^2$
or $m_\pi$.

This procedure yielded $M_\Omega r_1=2.666(41)(26)(+10-35)$, where
the errors are statistical, uncertainty in the lattice strange
quark mass, and a crude estimate of possible higher order terms
in the chiral and continuum extrapolations.  To get the $\Omega^-$ mass in 
physical units we use $r_1=0.317(7)$ fm, fixed from 
$\Upsilon$ splittings\cite{UPSILON}.


\begin{figure}[htb]
\scalebox{0.40}{\includegraphics{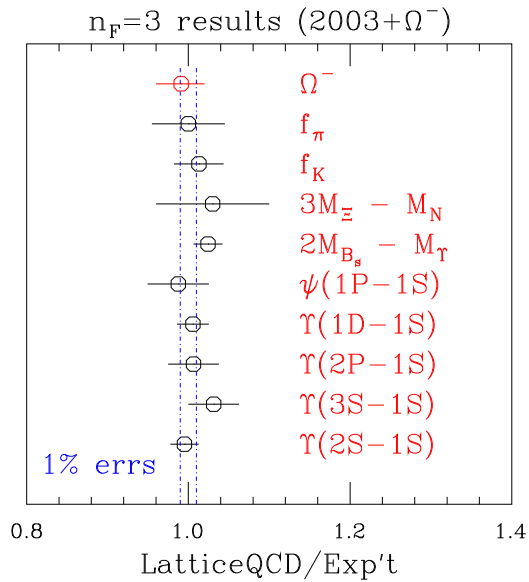}}
\vspace{-0.4in}
\caption{
Ratios of lattice results to physical values.  From Ref.~\protect\cite{PRL},
with preliminary $\Omega^-$ point added.
}
\label{new_pts_fig}
\end{figure}

Finally, to display the agreement with the experimental $\Omega^-$ mass
we plot in Fig.~\ref{new_pts_fig} the ratio of this lattice determination to the experimental
value on a plot from Ref.~\cite{PRL} with the ratios of several
other quantities to their experimental values.
Alternatively, if we require agreement of the $\Omega^-$ mass with experiment to
fix the lattice strange quark mass,
we find a result consistent with Ref.~\cite{QUARKMASS04} but with larger
statistical errors.


\section*{Acknowledgments}
This work was supported by PPARC, the DOE and the NSF.  We thank Urs Heller for
comments.
Computations were done at ORNL, NCSA, NERSC and the Univ. of Arizona.

\end{document}